\documentclass[twocolumn]{revtex4-1}
\usepackage{amsmath}
\usepackage{amsfonts}
\usepackage{amssymb}
\usepackage{dcolumn}
\usepackage{bm}
\usepackage[dvips]{graphicx}

\begin{document}

\title{The Loschmidt Echo as a robust decoherence quantifier for many-body systems}
\author{Pablo R. Zangara}
\author{Axel D. Dente}
\author{Patricia R. Levstein}
\author{Horacio M. Pastawski}
\email{horacio@famaf.unc.edu.ar}

\affiliation{Instituto de F\'{i}sica Enrique Gaviola (CONICET) and Facultad
de Matem\'{a}tica, Astronom\'{i}a y F\'{i}sica, Universidad Nacional
de C\'{o}rdoba, 5000, C\'{o}rdoba, Argentina}

\begin{abstract}
We employ the Loschmidt Echo, i.e. the signal recovered after the reversal of
an evolution, to identify and quantify the processes contributing to
decoherence. This procedure, which has been extensively used in single
particle physics, is here employed in a spin ladder. The isolated chains have
$1/2$ spins with $XY$ interaction and their excitations would sustain a
one-body like propagation. One of them constitutes the controlled system
$\mathcal{S}$ whose reversible dynamics is degraded by the weak coupling with
the uncontrolled second chain, i.e. the environment $\mathcal{E}$. The
perturbative $\mathcal{SE}$ coupling is swept through arbitrary combinations
of $XY$ and Ising like interactions, that contain the standard Heisenberg and
dipolar ones. Different time regimes are identified for the Loschmidt Echo
dynamics in this perturbative configuration. In particular, the exponential
decay scales as a Fermi golden rule, where the contributions of the different
$\mathcal{SE}$ terms are individually evaluated and analyzed. \ Comparisons
with previous analytical and numerical evaluations of decoherence based on the
attenuation of specific interferences,\textbf{\ }show that the Loschmidt Echo
is an advantageous decoherence quantifier at any time, regardless of the
$\mathcal{S}$ internal dynamics.

\end{abstract}

\pacs{03.67.Hk, 03.65.Yz, 75.10.Pq}
\maketitle

\section{INTRODUCTION}

The physical realization of Quantum Information Processing (QIP)
\cite{vicenzo}, requires a precise control of quantum dynamics. The coherent
manipulation of many-body systems plays a crucial role for several QIP-related
implementations, such as spintronic devices \cite{flatte07}, optical lattices
\cite{bloch1,bloch2}, superconducting circuits \cite{devoret2010} and
nitrogen-vacancy centers in diamond \cite{nv+MRI-lukin}. Experiments with spin
arrays in Nuclear Magnetic Resonance (NMR)
\cite{patricia98,usaj-physicaA,suterprl2004}, have shown that many-body
dynamics conspires against quantum control. Moreover, once the system interacts
with an environment, control becomes even more difficult due to information
leakage. This degradation of the system's coherent dynamics, called
decoherence, is subject of deep theoretical and experimental investigation
\cite{zurekpolonica}, as it remains the key obstacle for QIP.
Indeed, quantum error correction protocols
\cite{error-correction-1,error-correction-2} can restore quantum information
provided that they operate above a certain threshold. Achieving this limit is
often a task for dynamical decoupling techniques
\cite{violaDD,d-decoupling-cory,alvarezDD}. However, specific implementations
require a precise knowledge of the nature of the decoherence processes
\cite{laflamme-nature2010}. It is the purpose of this paper to contribute to a
better characterization of the role of a spin environment on the decoherence
of a many-spin system.


At least for short distance communications, spin chains can be used to
transfer information \cite{Bose2}. In fact, several selective polarization
techniques have been developed in NMR experiments to set up an initial local
excitation in one edge of a spin chain and transfer it to the other edge by
means of an effective $XY$ Hamiltonian (i.e. $S_{i}^{+}S_{j}^{-}+S_{i}%
^{-}S_{j}^{+}$ or polarization conserving "flip-flop" processes)
\cite{madi-ernst1997,cory-PRA07}. Additionally, Multiple Quantum Coherence
spectroscopy has allowed the study of quasi-one-dimensional spin systems under
the influence of spin environments \cite{cory-PRA09,elenaMQC}. Here,
the Double Quantum Hamiltonian (i.e. the $S_{i}^{+}S_{j}^{+}+S_{i}^{-}%
S_{j}^{-}$ processes), can be mapped to an $XY$ Hamiltonian allowing the
design and control of the excitation transfer in a broader family of
solid-state spin structures
\cite{feldman1997,feldman2000,feldman2005,cory-PRA07,cory2011}. Thus, a deep
knowledge of decoherence in such 1-D systems is crucial to improve the degree
of control available for NMR-based state transfer protocols
\cite{frydman,viola}.

A natural way to quantify the decoherence time $\tau_{\phi}$\ is through the
degradation of interferences. This requires the identification of specific
coherence \textquotedblleft witnesses\textquotedblright, such as excitations
in the local polarization. Particularly useful are the
reflections in the boundaries that can be observed as well defined Mesoscopic
Echoes (ME) \cite{mesoECO-theory,mesoECO-exp,altshuler94}. Recently, the ME
intensity has been used to quantify decoherence of spins arranged in a ladder
topology \cite{Alvarez-ME}. Alternatively, the evaluation of $\tau_{\phi}$ can
be performed by a time reversal procedure, the Loschmidt Echo (LE)
\cite{jalabert-hmp}, where one evaluates the reversibility of the system's
dynamics in the presence of an uncontrolled environment. The LE can be accessed
experimentally in many situations, such as spin systems
\cite{usaj-physicaA,patricia98,karina09}, confined atoms \cite{davidson} and
microwave excitations \cite{gorinMicroWave}. Besides, it has become a
standard way to quantify decoherence, stability and complexity in dynamical
processes, in several physical situations \cite{Jacquod,prosen,*loschScholarpedia}.

In the present article, we address the controllability of a spin chain
($\mathcal{S}$) in the presence of a spin environment ($\mathcal{E}$) by
performing a quantitative study of the LE. The LE degradation characterizes
the decoherence due to the perturbation of $\mathcal{E}$ on the otherwise
simple dynamics of $\mathcal{S}$. Indeed, the many-body nature of the
$\mathcal{S}$-$\mathcal{E}$ interaction yields a very rich behavior in the
dynamical regimes of decoherence: a short time quadratic decay, an exponential
regime, and a saturation plateau are identified by our numerical approach. In
particular, we perform a detailed analysis of the LE exponential decay,
addressing how the rates scale with a Fermi golden rule (FGR). Additionally,
since for weak perturbations the LE of the local excitation can be seen as a
Survival Probability (SP), the numerical results are compared to previous
analytical predictions for that magnitude \cite{Flambaum2001}.

In the next section, we describe the spin problem and summarize its
theoretical background. We introduce the spin autocorrelation function and
describe a single particle analogy which underlies further analysis. We also
discuss the local version of the Loschmidt Echo and present its definition in
terms of the local polarization, which is the usual experimental observable.
In Section \ref{Sec: Deco Eval LE}, we present the numerical study of the LE
for some of the different physically relevant parameters. We consider $\mathcal{S}$-$\mathcal{E}$
interactions which are weak
compared with those determining the $\mathcal{E}$ dynamics. We also address different
anisotropies of the $\mathcal{S}$-$\mathcal{E}$ interaction which ranges from
pure $XY$ (planar) to truncated dipolar cases, going through the Heisenberg
(isotropic) interaction. In Section \ref{Sec: Deco Rate Analysis} we analyze
the obtained results. First we consider the transition from the short time quadratic decay to the
exponential regime in analogy to what is known for the SP. Then, we focus on
the exponential regime to show that the Fermi golden rule in the present spin
problem has independent contributions arising from each specific process in
the $\mathcal{S}$-$\mathcal{E}$ interaction. We then compare these rates with
a previous evaluation based on the contrast of specific interferences (ME
attenuation). In the last section, we conclude that since the LE filters the
internal dynamics of the system, it provides a reliable and continuous access to all time
regimes. Thus, the LE compares favorably with the evaluation of decoherence
based on interference contrast.

\section{QUANTUM DYNAMICS IN SPIN SYSTEMS\label{Sec: QD of Spin Chain}}

\subsection{THE MODELS}

\begin{figure}
    \centering
     \includegraphics[width=0.5\textwidth]{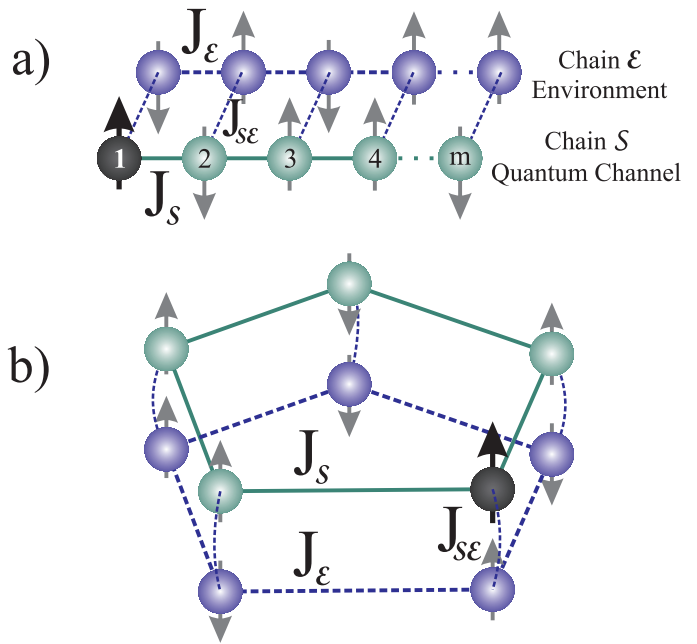}\\
      \caption{(Color Online). The spin system. (\textbf{a}) Open boundary
conditions. (\textbf{b}) Closed boundary conditions (ring-like). Continuous
(green) connections represent interactions that can be inverted to obtain the
Loschmidt echo. Dash (blue) lines represent non-controllable interactions. The
first spin (black circle) is initially polarized and the rest of the spins are
in one of the high temperature configurations.}%
    \label{Fig: system1}%
\end{figure}

The spin models analyzed in this article are schematized in Fig.
\ref{Fig: system1}. In the first one, the system $\mathcal{S}$ is an $m$-spin
chain (Fig. \ref{Fig: system1}-\textbf{a}), which could constitute a quantum
channel. It interacts with a second chain $\mathcal{E}$, that stands for the
\textquotedblleft environment\textquotedblright\ which perturbs the dynamics
of $\mathcal{S}$. The second model (Fig. \ref{Fig: system1}-\textbf{b}) is
obtained from the first one by imposing a periodic boundary condition that
transforms chains into rings.%

For both models, the spin Hamiltonian is given by:%

\begin{equation}
\hat{H}_{total}^{{}}=\hat{H}_{\mathcal{S}}^{{}}\otimes \hat{I}_{\mathcal{E}%
}+\hat{I}_{\mathcal{S}}\otimes\hat{H}_{\mathcal{E}}^{{}}+\hat{V}_{\mathcal{SE}}^{{}%
},\label{H1-1}%
\end{equation}
where the first and second terms represent the system and the environment
Hamiltonians respectively, and the third one is the interaction between them.
In order to simplify notation, we write just $\hat{H}_{\mathcal{S}}^{{}}$ and
$\hat{H}_{\mathcal{E}}^{{}}$\ instead of the tensor product with their respective
$\hat{I}_{\mathcal{E}}^{{}}$ and $\hat{I}_{\mathcal{S}}^{{}}$ identities.
For both $\mathcal{\nu}=\mathcal{S}$ or $\mathcal{E}$, we use an
effective \textquotedblleft planar\textquotedblright\ or $XY$ Hamiltonian
\cite{madi-ernst1997}, that describes the homogeneous flip-flop interaction
between nearest neighbor spins. In the model of Fig. \ref{Fig: system1}%
-\textbf{a}, i.e. the chain:%

\begin{align}
\hat{H}_{\nu}^{{}} &  =\sum_{n=1}^{m-1}J_{\nu}^{{}}(\hat{S}_{\mathcal{\nu
},n+1}^{x}\hat{S}_{\mathcal{\nu},n}^{x}+\hat{S}_{\mathcal{\nu},n+1}^{y}\hat
{S}_{\mathcal{\nu},n}^{y})\nonumber\\
&  =\sum_{n=1}^{m-1}\frac{J_{\nu}^{{}}}{2}(\hat{S}_{\mathcal{\nu},n+1}^{+}%
\hat{S}_{\mathcal{\nu},n}^{-}+\hat{S}_{\mathcal{\nu},n+1}^{-}\hat
{S}_{\mathcal{\nu},n}^{+}).\label{H2-2}%
\end{align}
Here $\hat{S}_{\mathcal{\nu},n}^{x}$ and $\hat{S}_{\mathcal{\nu},n}^{y}$ are
the $x$ and $y$ components of the spin operator in the $n$-th site in the
$\nu$ chain respectively, while $\hat{S}_{\mathcal{\nu},n}^{+}$ and $\hat
{S}_{\mathcal{\nu},n}^{-}$ are the raising and lowering operators. Again, the
abbreviated notation for any spin operator must be understood in the form
$\hat{S}_{\mathcal{S},n}^{z}=\hat{I}_{1}\otimes\ldots\otimes\hat{S}_{n}^{z}%
\otimes\ldots\otimes \hat{I}_{m}\otimes \hat{I}_{m+1}\otimes\ldots\otimes \hat{I}_{2m}$. If one
wants to consider the ring model, an extra $XY$ coupling appears between the
$1$-st and $m$-th spins.

The interchain coupling is:%

\begin{align}
\hat{V}_{\mathcal{SE}}^{{}} &  =\sum_{n=1}^{m}J_{\mathcal{SE}}^{{}}%
[2\alpha\hat{S}_{\mathcal{S},n}^{z}\hat{S}_{\mathcal{E},n}^{z}-(\hat
{S}_{\mathcal{S},n}^{x}\hat{S}_{\mathcal{E},n}^{x}+\hat{S}_{\mathcal{S},n}%
^{y}\hat{S}_{\mathcal{E},n}^{y})]\label{H3}\\
&  =\sum_{n=1}^{m}J_{\mathcal{SE}}^{{}}[2\alpha\hat{S}_{\mathcal{S},n}^{z}%
\hat{S}_{\mathcal{E},n}^{z}-\frac{1}{2}(\hat{S}_{\mathcal{S},n}^{+}\hat
{S}_{\mathcal{E},n}^{-}+\hat{S}_{\mathcal{S},n}^{-}\hat{S}_{\mathcal{E},n}%
^{+})],\label{H3-3}%
\end{align}
where the first term is an Ising interaction. The $\alpha$ parameter
determines the anisotropy of the coupling. This encompass the typical magnetic
resonance scenarios: the $XY$ (planar)
interaction \cite{madi-ernst1997}, represented by $\alpha=0$; the
Heisenberg (isotropic) interaction \cite{LevPasCalvo1989}, by $\alpha=-\frac{1}{2}$; and
the truncated dipolar coupling \cite{Ernst92-PolarizEcho} corresponds to
$\alpha=1$. In order to extend and systematize our analysis we also consider
several other values for $\alpha$. It is important to notice that for any
finite $\alpha$ the $\mathcal{S}$-$\mathcal{E}$ interaction has always an $XY$
component. This allows a polarization exchange which, in a Fermionic
representation, can be seen as a \textquotedblleft single-particle
tunnelling\textquotedblright\ \cite{danieli-CPL2004}. In such a picture, the
Ising term corresponds to a nearest neighbour Hubbard term which is a two-body interaction.

It is crucial to stress that the real constants $J_{\mathcal{S}}^{{}}$,
$J_{\mathcal{E}}^{{}}$ and $J_{\mathcal{SE}}^{{}}$ determine the relevant time
scales of the whole dynamics. As introduced above, the first two give the
homogeneous $XY$ coupling within $\mathcal{S}$ and $\mathcal{E}$ respectively,
while $J_{\mathcal{SE}}^{{}}$ stands for the interchain coupling. To ensure a
smooth degradation of the $\mathcal{S}$ coherent dynamics, we set
$J_{\mathcal{SE}}^{{}}$ in the weak coupling limit, i.e. $J_{\mathcal{SE}}%
^{{}}\ll$ $J_{\mathcal{S}}^{{}}=J_{\mathcal{E}}^{{}}$.

\subsection{Basic features of Spin Dynamics}

A natural question for the spin models introduced above is how to
quantify the decoherence of $\mathcal{S}$\ in the presence of $\mathcal{E}$. This
means that one has to deal with a composite (bipartite) Hilbert space
$\mathcal{S\otimes E}$, and trace out the\textbf{\ }$\mathcal{E}$-degrees of
freedom whenever necessary. A standard strategy relies on finding an
appropriate decoherence rate $1/\tau_{\phi}$ through the quantification of the
attenuation of system's specific interferences. As in the experiments
\cite{Ernst92-PolarizEcho, mesoECO-exp}, one starts evaluating the evolution
of an injected local polarization through the spin autocorrelation function
\cite{Abragam-Book,LevPasCalvo1989}:%

\begin{equation}
P_{1,1}^{{}}(t)=\frac{\left\langle \Psi_{eq}^{{}}\right\vert \hat
{S}_{\mathcal{S},1}^{z}(t)\hat{S}_{\mathcal{S},1}^{z}(0)\left\vert \Psi
_{eq}^{{}}\right\rangle }{\left\langle \Psi_{eq}^{{}}\right\vert \hat
{S}_{\mathcal{S},1}^{z}(0)\hat{S}_{\mathcal{S},1}^{z}(0)\left\vert \Psi
_{eq}^{{}}\right\rangle }.\label{autocorrelation}%
\end{equation}
This function gives the local polarization at time $t$ along the $z$ direction
in site $1$ provided that at time $t=0$ the system was in its thermal
equilibrium state plus a local excitation in site $1$. Here, the spin operator
in the Heisenberg representation is given by $\hat{S}_{\mathcal{S},1}%
^{z}(t)=e^{\mathrm{i}\hat{H}_{total}^{{}}t/\hbar}\hat{S}_{\mathcal{S},1}%
^{z}e^{-\mathrm{i}\hat{H}_{total}^{{}}t/\hbar}$. The many-body state
$\left\vert \Psi_{eq}^{{}}\right\rangle $ corresponding to a high temperature
thermal equilibrium represents a mixture of all states with amplitudes
satisfying the appropriate statistical weights and random phases. Then, the
initial non-equilibrium local excitation $\left\vert \Psi_{ne}^{{}%
}\right\rangle $ can be defined in terms of the computational (Ising) basis
components that have the $1^{\mathrm{st}}$ spin up as:
\begin{equation}
\left\vert \Psi_{ne}^{{}}\right\rangle =\frac{\hat{S}_{\mathcal{S},1}%
^{+}\left\vert \Psi_{eq}^{{}}\right\rangle }{\left\vert \left\langle \Psi
_{eq}^{{}}\right\vert \hat{S}_{\mathcal{S},1}^{-}\hat{S}_{\mathcal{S},1}%
^{+}\left\vert \Psi_{eq}^{{}}\right\rangle \right\vert ^{1/2}}=\sum_{r}%
c_{r}\left\vert \Psi_{r}^{{}}\right\rangle ,\label{exitacion-inicial}%
\end{equation}
see Appendix
\ref{Sec: Apendice2} for further details. Here, each contribution $\left\vert \Psi_{r}^{{}}\right\rangle $ to the
locally polarized initial state can be written as:%

\begin{equation}
\left\vert \Psi_{r}^{{}}\right\rangle =\left\vert \uparrow_{1}\right\rangle
\otimes\left\vert \beta_{r}^{{}}\right\rangle ,\label{contrib-r}%
\end{equation}
where the basis for the remaining $2m-1$ spins is%
\begin{equation}
\left\vert \beta_{r}\right\rangle =\left\vert s_{2}\right\rangle
\otimes\left\vert s_{3}\right\rangle \otimes\left\vert s_{4}\right\rangle
\otimes...\otimes\left\vert s_{2m}\right\rangle \ \text{\ with }\left\vert
s_{k}\right\rangle \in\left\{  \left\vert \uparrow\right\rangle ,\left\vert
\downarrow\right\rangle \right\}  .\label{beta-r}%
\end{equation}

Since in the regime of NMR spin dynamics, the thermal energy $k_{B}T$
is much higher than any other energy scale of the system
\cite{Slichter-Book,Abragam-Book}, all statistical weights result identical,
i.e. $\left\vert c_{r}\right\vert =1/\sqrt{2^{2m-1}}.$

It is useful to analyze the autocorrelation $P_{1,1}^{{}}(t)$. For
each contribution $\left\vert \Psi_{r}\right\rangle $ to the initial state,
the evolved wave function at time $t$, is%
\begin{equation}
\left\vert \Psi_{r}(t)\right\rangle =\exp\{-\tfrac{\mathrm{i}}{\hbar}\hat
{H}_{total}^{{}}t\}\left\vert \Psi_{r}\right\rangle .\label{evol-r}%
\end{equation}
Then, the probability of finding the first spin up-polarized is:%

\begin{equation}
P_{1,1}^{\left[  r\right]  }(t)=\sum_{j}\left\vert (\left\langle \uparrow
_{1}\right\vert \otimes\left\langle \beta_{j}\right\vert )\left\vert \Psi
_{r}(t)\right\rangle \right\vert ^{2},\label{p11-r}%
\end{equation}
where the sum runs over the 2$^{2m-1}$ configurations of the $2m-1$ remaining
spins. Notice that the sum over the $j$-index in Eq. \ref{p11-r}, means that
we are performing a trace not only over $\mathcal{E}$, but also over the spins
in $\mathcal{S}$\ other than the first one. After the summation over all
contributions $\left\vert \Psi_{r}\right\rangle $ to the initial state, and
expressing the result as a local polarization \cite{mesoECO-theory}, we obtain:%

\begin{equation}
P_{1,1}(t)=\bigg[\sum_{r=1}^{2^{2m-1}}\left\vert c_{r}\right\vert ^{2}P_{1,1}^{
\left[  r\right]  }(t)-\frac{1}{2}\bigg]\times2.\label{p11}%
\end{equation}
A computation of the time dependent local observable in Eq. \ref{autocorrelation} reduces to
Eq. \ref{p11} which in turn, requires evolving
each of the $2^{2m-1}$ pure states to evaluate the ensemble averaged
observables. This is implemented using a Trotter decomposition \cite{trotter}
assisted by an algorithm that exploits quantum parallelism
\cite{Alv-parallelism} (See appendix \ref{Sec: Apendice}).

\subsection{The single-particle picture and Mesoscopic Echoes}

The obvious complexity of the many spin dynamics might hinder some simple
interference phenomena that can be taken as hallmarks. Under certain
experimentally achievable conditions (e.g. $XY$ interaction, 1D topology and
negligible $\mathcal{S}$-$\mathcal{E}$ interaction), the autocorrelation
$P_{1,1}(t)$ becomes a simple one-body magnitude. Indeed, once the initial
excitation is created, the first physical picture about its evolution may be
obtained from the Wigner-Jordan spin-fermion mapping \cite{Lieb-Mattis1961}.
The point here is to trace over the $\mathcal{E}$-degrees of freedom from the
beginning and focus on a single quantum spin chain. Accordingly, an
isolated $m$ spin chain with $XY$ interaction, where $N$ of them are up, is
mapped to a chain with $N$ non-interacting fermions. Thus, a local
polarization excitation has the same dynamics as a single fermion wave-packet
in a tight-binding linear chain \cite{danieli-CPL2004,danieli-CPL2005}.
The observed autocorrelation function in the limit of infinite
temperature is precisely described by the evolution of a single spin up in a
chain of down spins:%

\begin{equation}
\left\vert \Psi_{1}\right\rangle :=\left\vert \uparrow_{1}\right\rangle
\otimes\left\vert \downarrow_{2}\right\rangle \otimes\left\vert \downarrow
_{3}\right\rangle \otimes...\otimes\left\vert \downarrow_{m}\right\rangle.
\label{psi-trazado}%
\end{equation}

This is a one-body wave function, defined on a subspace of $\mathcal{S}$
where the total spin projection is $\frac{m}{2}-1$, and the observable is
evaluated as:%
\begin{equation}
P_{1,1}^{\left[  r=1\right]  }(t)=\left\vert \left\langle \Psi_{1}\right\vert
\exp[-\mathrm{i}\hat{H}_{\mathcal{S}}^{{}}t/\hbar]\left\vert \Psi
_{1}\right\rangle \right\vert ^{2}.\label{1BODY-local-polarization}%
\end{equation}
Since this is a finite size system one should expect revivals of the initial
polarization. Such revivals are called Mesoscopic Echoes (ME)
\cite{altshuler94,mesoECO-theory} and they appear when constructive
interferences manifest at the Heisenberg time $t_{H}\sim\hbar/\Delta$, with
$\Delta$ being the typical mean energy level spacing. In a spin chain with
$XY$ interaction one may safely use $\Delta\simeq J_{\mathcal{E}}^{{}}/m$. As
long as such one-body picture remains approximately valid for a linear chain
weakly coupled to the environment these interferences show up
experimentally \cite{mesoECO-exp,madi-ernst1997,Khitrin}.

A weak coupling to the spin bath $\mathcal{E}$ results in a progressive
attenuation of the ME, which has been used to quantify environmentally induced
decoherence \cite{Alvarez-ME}. This attenuation is understood as a Fermi
golden rule (FGR), which describes an \textquotedblleft
irreversible\textquotedblright\ decay of a pure state in $\mathcal{S}$\ into
collective $\mathcal{S\otimes E}$\ states. In fact, the validity of the FGR
requires here the breakdown of degeneracies, i.e. the whole $\mathcal{S}%
\otimes\mathcal{E}$ must behave as a fully many-body system. We shall return
to this point below.

\subsection{Loschmidt Echo}

Let us now explain the essence of the protocol that uses the Loschmidt Echo to
quantify decoherence in a spin system using a local spin as an observable
\cite{patricia98}. Our strategy relies on the controllability of the chain
$\mathcal{S}$, whose Hamiltonian's sign can be switched at will as it is often
the case in NMR. The reversibility of the dynamics within the chain is
perturbed by the interaction with the non-controlled spin chain $\mathcal{E}$.
There are two stages in the evolution. First, a spin excitation is created and
the whole spin set evolves according to the Hamiltonian of Eq. \ref{H1-1},
during a time $t_{R}$. At that time, the internal interactions within
$\mathcal{S}$ are reversed (i.e. $\hat{H}_{\mathcal{S}}^{{}}$ is replaced by
$-\hat{H}_{\mathcal{S}}^{{}}$ during a second $t_{R}$-period). However,
neither the $\mathcal{S}$-$\mathcal{E}$ coupling nor the interactions within
chain $\mathcal{E}$ are reversed, leading to a non-reversed perturbation,
\begin{equation}
\hat{\Sigma}=\hat{H}_{\mathcal{E}}^{{}}+\hat{V}_{\mathcal{SE}}^{{}%
},\label{sigma}%
\end{equation}
acting in \textit{both} periods. Thus, in analogy with Eq.
\ref{autocorrelation}, we define the observable Loschmidt echo as the
recovered local polarization:%

\begin{equation}
M_{LE}(2t_{R})=\frac{\left\langle \Psi_{eq}^{{}}\right\vert \hat
{S}_{\mathcal{S},1}^{z}(2t_{R})\hat{S}_{\mathcal{S},1}^{z}(0)\left\vert
\Psi_{eq}^{{}}\right\rangle }{\left\langle \Psi_{eq}^{{}}\right\vert \hat
{S}_{\mathcal{S},1}^{z}(0)\hat{S}_{\mathcal{S},1}^{z}(0)\left\vert \Psi
_{eq}^{{}}\right\rangle }.\label{loca-Many-Body Loschmidt}%
\end{equation}
The spin operators, expressed in the Heisenberg representation, are now:
\begin{equation}
\hat{S}_{\mathcal{\nu},1}^{z}(2t_{R})=e^{\frac{\mathrm{i}}{\hbar}(-\hat
{H}_{\mathcal{S}}^{{}}+\hat{\Sigma})t_{R}}e^{\frac{\mathrm{i}}{\hbar}(\hat
{H}_{\mathcal{S}}^{{}}+\hat{\Sigma})t_{R}}\hat{S}_{\mathcal{\nu},1}%
^{z}e^{-\frac{\mathrm{i}}{\hbar}(\hat{H}_{\mathcal{S}}^{{}}+\hat{\Sigma}%
)t_{R}}e^{-\frac{\mathrm{i}}{\hbar}(-\hat{H}_{\mathcal{S}}^{{}}+\hat{\Sigma
})t_{R}}.\label{S-heisenberg}%
\end{equation}
The computation of the Loschmidt Echo in Eq. \ref{loca-Many-Body Loschmidt}
proceeds as we did above for the forward dynamics. In a system with a general
interaction, it requires a full many-body evolution.

As before, the initial excitation is described in terms of the Ising basis by
Eq. \ref{exitacion-inicial}. For each contribution $\left\vert \Psi
_{r}\right\rangle $ to the initial state, the
resulting wave function at time $t=2t_{R}$, after the whole time-reversal
procedure, is%
\begin{equation}
\left\vert \Psi_{r}(2t_{R})\right\rangle =\exp\{-\tfrac{\mathrm{i}}{\hbar
}(-\hat{H}_{\mathcal{S}}^{{}}+\hat{\Sigma})t_{R}\}\exp\{-\tfrac{\mathrm{i}%
}{\hbar}(\hat{H}_{\mathcal{S}}^{{}}+\hat{\Sigma})t_{R}\}\left\vert \Psi
_{r}\right\rangle .\label{psi-revertido}%
\end{equation}
In analogy with the discussion above, the probability of finding the first
spin up-polarized is:%
\begin{equation}
M_{1,1}^{\left[  r\right]  }(t)=\sum_{j}\left\vert (\left\langle
\uparrow_{1}\right\vert \otimes\left\langle \beta_{j}\right\vert )\left\vert
\Psi_{r}(t)\right\rangle \right\vert ^{2},\label{maglocal}%
\end{equation}
where the sum runs over the 2$^{2m-1}$ configurations of the $2m-1$ remaining
spins.\ Again, the sum over $j$ means a trace over all the spins of the
environment and the system except the first one. After the summation (average)
over all contributions $\left\vert \Psi_{r}\right\rangle $ to the initial
state, and expressing the result as a local polarization, we compute the local
Loschmidt Echo as:%
\begin{equation}
M_{LE}(t)=\bigg[\sum_{r=1}^{2^{2m-1}}\left\vert c_{r}\right\vert ^{2}M_{1,1}^{
\left[  r\right]  }(t)-\frac{1}{2}\bigg]\times2.\label{maglocal2}%
\end{equation}
Here again, the statistical weights $\left\vert c_{r}\right\vert ^{2}$ are the
inverse of $2^{2m-1},$ the number of initial states \ in the ensemble that
satisfy the \textquotedblleft$1^{\mathrm{st}}$\textit{\ spin up polarized}%
\textquotedblright\ condition. Notice that, except for the fact that the
evolution operator contains a partially reversed dynamics, this quantity
refers to the same physical observable as Eq. \ref{p11}.

\subsection{Connection to previous works}

It would be useful to make a connection between the observable just described
and the usual definition of the Loschmidt Echo \cite{jalabert-hmp}, which was
extensively studied in the field of quantum chaos and quantum information
\cite{Jacquod,prosen,*loschScholarpedia}. With this purpose, let us consider the particular case
of Fig. \ref{Fig: system1} where $\hat{H}_{\mathcal{S}}^{{}}$ describes a
chain with $XY$ interactions, while $\hat{H}_{\mathcal{E}}^{{}}$ describes a
chain $\mathcal{E}$ that remains quenched in a random configuration and
$\hat{V}_{\mathcal{SE}}^{{}}$ is restricted to an Ising interaction. Under
these assumptions, the time-reversed dynamics reduces to that of a single
spin up in an oriented chain defined in Eq. \ref{psi-trazado}. Accordingly,
$\hat{\Sigma}$ becomes an Hermitian self-energy operator $\hat{\Sigma
}_{\mathcal{S}}$ acting on the $\mathcal{S}$ Hilbert space. Indeed,
$\hat{\Sigma}_{\mathcal{S}}$ represents a set of non-reversed random energy
shifts produced by the Ising interaction with the static environmental spins.
In the independent fermion picture, $\hat{\Sigma}_{\mathcal{S}}$ yields a
\textquotedblleft random potential\textquotedblright for each specific
configuration of the environment, i.e. the the binary alloy variant
of an Anderson's disorder \cite{AndersonRMP-1978}. It is now relevant to address the physical
meaning of the sum over the $2^{2m-1}$ indices $r$ of the observable in Eq. \ref{maglocal2}.
This sum performs a trace over system spins which evolve
but are not observed (crucial to recover a one-body dynamics), as well as
a trace over the environmental spins, which can be seen as an ensemble
average over quenched disorder configurations \cite{jalabert-hmp}. Thus, the
same procedure that enabled to reduce Eq. \ref{autocorrelation} to Eq.
\ref{1BODY-local-polarization}, transforms Eq. \ref{loca-Many-Body Loschmidt}
into the corresponding one-body Loschmidt Echo:%
\begin{widetext}
\begin{equation}
M_{LE}(2t_{R})=\left\vert \left\langle \Psi_{1}\right\vert \exp\{-\frac
{\mathrm{i}}{\hbar}(-\hat{H}_{\mathcal{S}}^{{}}+\hat{\Sigma}_{\mathcal{S}%
})t_{R}\}\exp\{-\frac{\mathrm{i}}{\hbar}(\hat{H}_{\mathcal{S}}^{{}}%
+\hat{\Sigma}_{\mathcal{S}})t_{R}\}\left\vert \Psi_{1}\right\rangle
\right\vert ^{2}.\label{simplifiedMBloschmidt}%
\end{equation}
\end{widetext}
Here, one can recognize the LE definition introduced in Ref.
\cite{jalabert-hmp} as the overlap of two wave functions evolving in presence
of a quenched disorder whose dynamics is not inverted.

The decomposition of the spin set into a controllable subset ($\mathcal{S}$)
and an uncontrollable one ($\mathcal{E}$) resembles the discussion of the
partial fidelity, called Boltzmann echo, analyzed in Ref.
\cite{Jacquod-prl2006} for a two-body problem. Analogously, the $2m$ spin
problem treated here verifies that: (i) a separation into two interacting
subsystems ($\mathcal{S}$ and $\mathcal{E}$) is performed, (ii) the initial state in
$\mathcal{S}$ is at least partially prepared (injected polarization), and at
the end a local measure is performed in the same site of injection, (iii) the
subsystem $\mathcal{E}$\ remains in the high temperature thermal equilibrium,
and (iv) the Hamiltonian in $\mathcal{S}$\ can be fully time-reversed, while
the Hamiltonian of $\mathcal{E}$\ and the $\mathcal{S}$-$\mathcal{E}%
$\ interaction remain uncontrolled. However, the main result of Ref.
\cite{Jacquod-prl2006}, that considers a \textit{chaotic} one-body system
coupled to a \textit{chaotic} one-body environment, can not be directly
compared with our study. Here, we focus on a system and an environment which
are both many-body systems that can be reduced to \textit{integrable} one-body
systems. In our case, we expect many-body chaos \cite{Bohigas1971} only as a
consequence of the $\mathcal{S}$-$\mathcal{E}$ interaction.

Many-body chaos has been the subject of much interest
\cite{Flambaum96,*Flambaum97,Shepe97,Shepe98}, which was recently renewed mainly in
connection to thermalization dynamics \cite{izrailev2012}. Within this
context, the attention has been centered on the study of spectral correlations
and the related survival probability (SP), i.e. the decay of an initial
excitation \cite{Flambaum2001}. Based on general considerations on the
strength function or local density of states (LDoS) \cite{Flambaum2000}, one
may predict a Gaussian (or quadratic) time decay governed by the second moment
of the perturbation $\sigma^{2}$, followed by an exponential regime whose rate is described
by the FGR:%
\begin{equation}
1/\tau_{FGR}=\frac{2\Gamma}{\hbar}=\frac{2\pi}{\hbar}\sigma^{2}N_{1}%
.\label{FGR-simple}%
\end{equation}
Here, $N_{1}^{{}}$ is the density of directly connected states (DDCS). While it might
seem elusive, this magnitude can be numerically evaluated with a Lanczos
algorithm \cite{ElenaPhysicaB}. It also accounts for the transition time from
the short-time quadratic regime to the exponential one. Accordingly, the
cross-over to the exponential regime is expected
\cite{Flambaum2001,ElenaBrazJ} to occur at $t_{s}\simeq(\tau_{FGR}\sigma^{2})^{-1}=(2\pi/\hbar) N_{1}$.
This  \textit{spreading time} characterizes the dissemination of an excitation within the environment. In the
many-body context it has been proposed an interpolation formula
\cite{Flambaum2001},%
\begin{equation}
P_{11}(t)=\exp[2\frac{\Gamma^{2}}{\sigma^{2}}-2\sqrt{\frac{\Gamma^{4}}%
{\sigma^{4}}-\Gamma^{2}t^{2}/\hbar^{2}}],\label{Gauss-FGR-interpolation}%
\end{equation}
which, to our knowledge, has not been checked in concrete systems.

Coming back to our model, when the chain $\mathcal{S}$\ is isolated, the
localized excitation propagates freely, and the LE would have a constant value
of $1$. However, once a weak $\hat{V}_{\mathcal{SE}}^{{}}$ interaction is turned
on, the excitation decays into the chain $\mathcal{E}$ and the local LE
degrades progressively with a law that would be closely described as a SP.
Thus, for long chains where the spectrum is a quasi-continuum, we expect that
the short and intermediate time-regimes of the LE would follow closely the
above discussion for the SP. In fact, we analyze the LE decay numerically and
study the rates testing the validity of a FGR description
\cite{Jacquod-Beenakker-2001,*Cucchietti-Lewenkopf-2006}. When applied to the
present many-body context it would look:%

\begin{equation}
\frac{1}{\tau_{\phi}}\simeq%
{\displaystyle\sum\limits_{\delta}}
\frac{2\pi}{\hbar}\left\vert V_{\mathcal{SE}}^{\delta}\right\vert
^{2}N_{1\delta},\label{fgr1}%
\end{equation}
where $\left\vert V_{\mathcal{SE}}^{\delta}\right\vert ^{2}\equiv\sigma^{2}$
is the local second moment of the process $\delta$ (e.g. Ising or $XY$)
contributing to the $\mathcal{S}$-$\mathcal{E}$ interaction and $N_{1\delta}$
represents some appropriate DDCS.

\section{QUANTIFYING DECOHERENCE THROUGH THE LOSCHMIDT ECHO }

\subsection{Numerical Results in Spin Chains\label{Sec: Deco Eval LE}}

\begin{figure}
    \centering
     \includegraphics[width=0.5\textwidth]{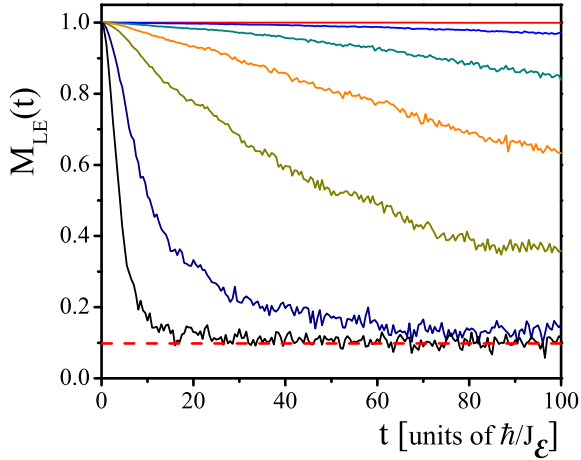}\\
      \caption{(Color Online). Time regimes of the local Loschmidt echo. Numerical
results for a ring of 5 spins weakly coupled to another identical ring by an
$XY$ interchain interaction ($\alpha=0$). The $M_{LE}$ is plotted as a
function of the total evolution time $t=2t_{R}$. Notice that the stronger the
interchain coupling $J_{\mathcal{SE}}^{{}}$, the faster the saturation regime
is reached (dashed line). From up to down, the different curves correspond to the coupling
parameters $J_{\mathcal{SE}}^{{}}$: $0.001,$ $0.01,$ $0.025,$ $0.05,$ $0.1,$
$0.25$ and $0.5 $, in units of $J_{\mathcal{E}}^{{}}$.}%
    \label{Fig: Short and Long Decay}%
\end{figure}

In this section we present the results obtained for $M_{LE}(t)$ in the spin
models represented in Fig. \ref{Fig: system1}. Even though our major concern
lies on the exponential decay described by the FGR, we can also identify the short
time quadratic decay and the saturation regime, as shown in Fig.
\ref{Fig: Short and Long Decay}. It is noticeable that the Loschmidt echo
yields results for a very wide range of parameters and times scales. This
feature contrasts with the study of interferences through the ME whose
observability restricts the quantification of decoherence.%

\begin{figure}
    \centering
     \includegraphics[width=0.5\textwidth]{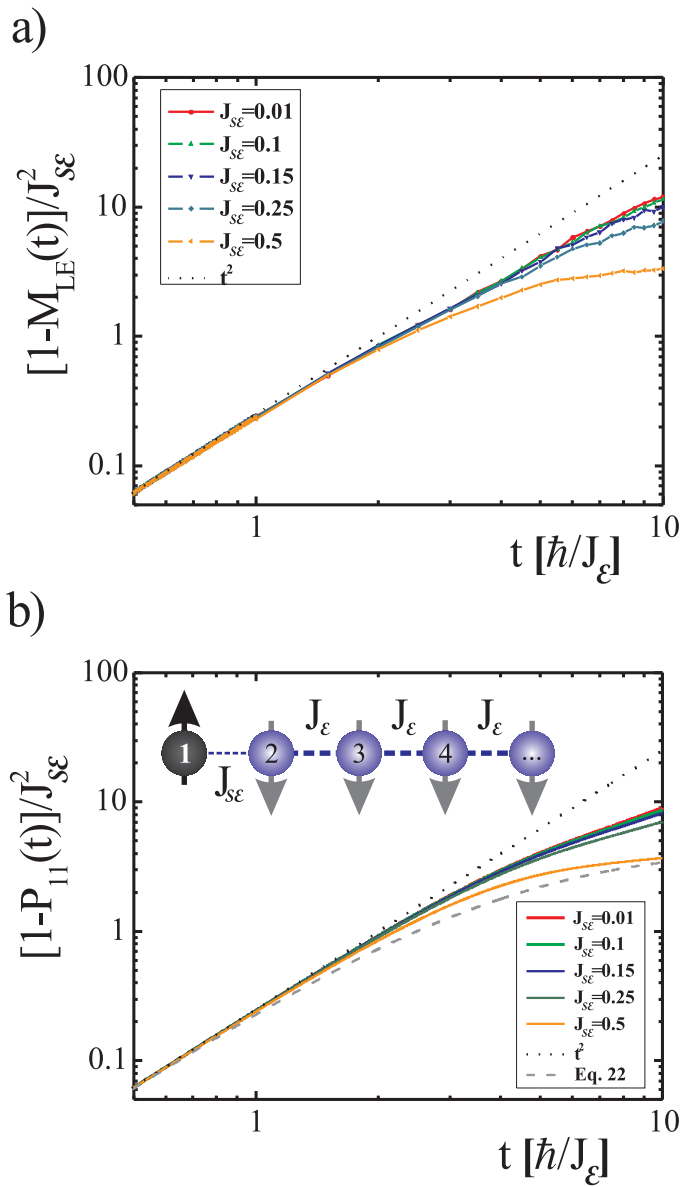}\\
      \caption{(Color online). \textbf{a})Short time behavior of the Loschmidt Echo. The dotted line
is the quadratic decay (Eq. \ref{Eq: mpe2}). \textbf{b})Short time behavior of the Survival Probability for an XY spin chain described by:\ $\hat
{H}_{total}=J_{\mathcal{SE}}^{{}}\left(  \hat{S}_{1}^{+}\hat{S}_{2}^{-}
+\hat{S}_{2}^{+}\hat{S}_{1}^{-}\right)  +J_{\mathcal{E}}^{{}}%
\sum_{n=2}^{\infty}\left(  \hat{S}_{n}^{+}\hat{S}_{n+1}^{-}+\hat{S}_{n+1}%
^{+}\hat{S}_{n}^{-}\right)  $, where the first spin is the system and the others constitute the environment. This system is known to satisfy the FGR \cite{elenacpl}. Notice the
similarity between (\textbf{a}) and (\textbf{b}) in departing from the
quadratic decay, which evinces the spreading time $t_{s}$. The dashed line
shows a plot of Eq. \ref{Gauss-FGR-interpolation} for $J_{\mathcal{SE}}^{{}%
}=0.5J_{\mathcal{E}}^{{}}$.}%
    \label{Fig: short-time-decay}%
\end{figure}

In Fig. \ref{Fig: short-time-decay}-\textbf{a} the short time dynamics is
compared with the expected quadratic decay (dotted line), which should appear
in any quantum evolution at short times. Indeed,\textbf{\ }the plot of
$\left(  1-M_{LE}\right)  /J_{\mathcal{SE}}^{2}$ (in log-log scale) as a
function of time, shows that $M_{LE}$ follows a quadratic function:%

\begin{equation}
M_{LE}(t)\simeq1-[\frac{J_{\mathcal{SE}}^{{}}}{2\hbar}]^{2}t^{2}%
.\label{Eq: mpe2}%
\end{equation}
This confirms that the short time decay is determined by $\sigma^{2}=(J_{\mathcal{SE}}^{{}}/2\hbar)^{2}%
$, the second moment of
the $\mathcal{SE}$ interaction. The agreement is observed until a time $t_{s}\simeq\hbar/J_{\mathcal{E}}$,
which verifies the prediction for $t_{s}$ in terms of the $\mathcal{E}$
dynamics. For comparison, we show in Fig. \ref{Fig: short-time-decay}%
-\textbf{b} the SP of an excitation in a single spin system $\mathcal{S}$ that
interacts with the edge of a spin chain $\mathcal{E}$ where it can spread
through pure $XY$ interactions. This model, shown in the inset, constitutes a
paradigm for the onset of the FGR, because the DDCS is precisely defined by
$N_{1}=1/J_{\mathcal{E}}$ and remains independent of $J_{\mathcal{SE}%
}^{{}}$ \cite{elenacpl}. In Fig. \ref{Fig: short-time-decay}-\textbf{b}, we
also show with a dashed line the interpolative expression in Eq.
\ref{Gauss-FGR-interpolation}, for the strongest coupling. This last
expression deviates faster from the quadratic decay than the SP in the
one-body chain.%

\begin{figure}
    \centering
     \includegraphics[width=0.5\textwidth]{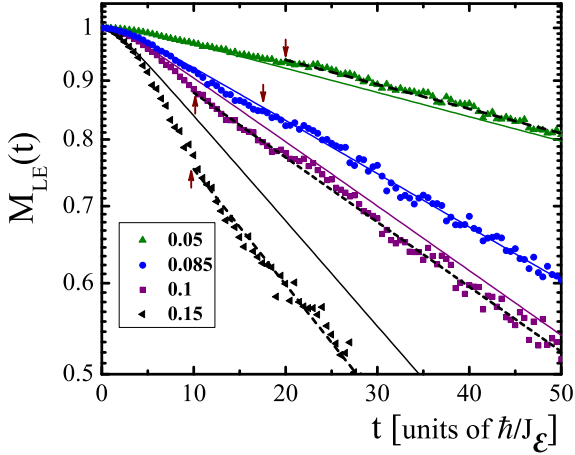}\\
      \caption{(Color online). The onset of the LE exponential decay. In the log
scale, the dashed lines are the ideal exponentials determined by the fitting,
show here for representative values. The delay in the entrance into the
exponential regime is indicated with arrows pointing the first data used in
the fitting. Continuous lines represent the interpolative Eq.
\ref{Gauss-FGR-interpolation}, used with the corresponding $\sigma^{2}$ and
$\Gamma$.}%
    \label{Fig: comparison}%
\end{figure}

The onset of the exponential regime is exemplified in Fig.
\ref{Fig: comparison}, for a few values of a pure XY
$\mathcal{SE}$ interaction. As a general tendency, we observe that the
asymptotic exponential decay only becomes well defined after some evolution
period. We indicate with an arrow the initial data that we use to fit these
rates. For comparison, the SP interpolative expression is plotted for the same decay rates. We notice
that \ for $J_{\mathcal{SE}}^{{}}/J_{\mathcal{E}}^{{}}\lesssim0.085$ the
interpolation starts to lie below the numerical results, and longer times
are required to define the asymptotic exponential.

For long times, the LE shows a saturation plateau at $1/10$ (see Fig.
\ref{Fig: Short and Long Decay}). Such observation is consistent
with the expectation that, within these coupling networks, the finite system
of interacting spins behaves ergodically under the Loschmidt Echo dynamics,
and thus the polarization spreads uniformly. At long times each site is
polarized by the amount $1/(2m)$\.
The larger the\ coupling $J_{\mathcal{SE}}^{{}}$, the faster the asymptotic
saturation is reached.

\begin{figure*}
[ptb]
    \centering
     \includegraphics[width=0.8\textwidth]{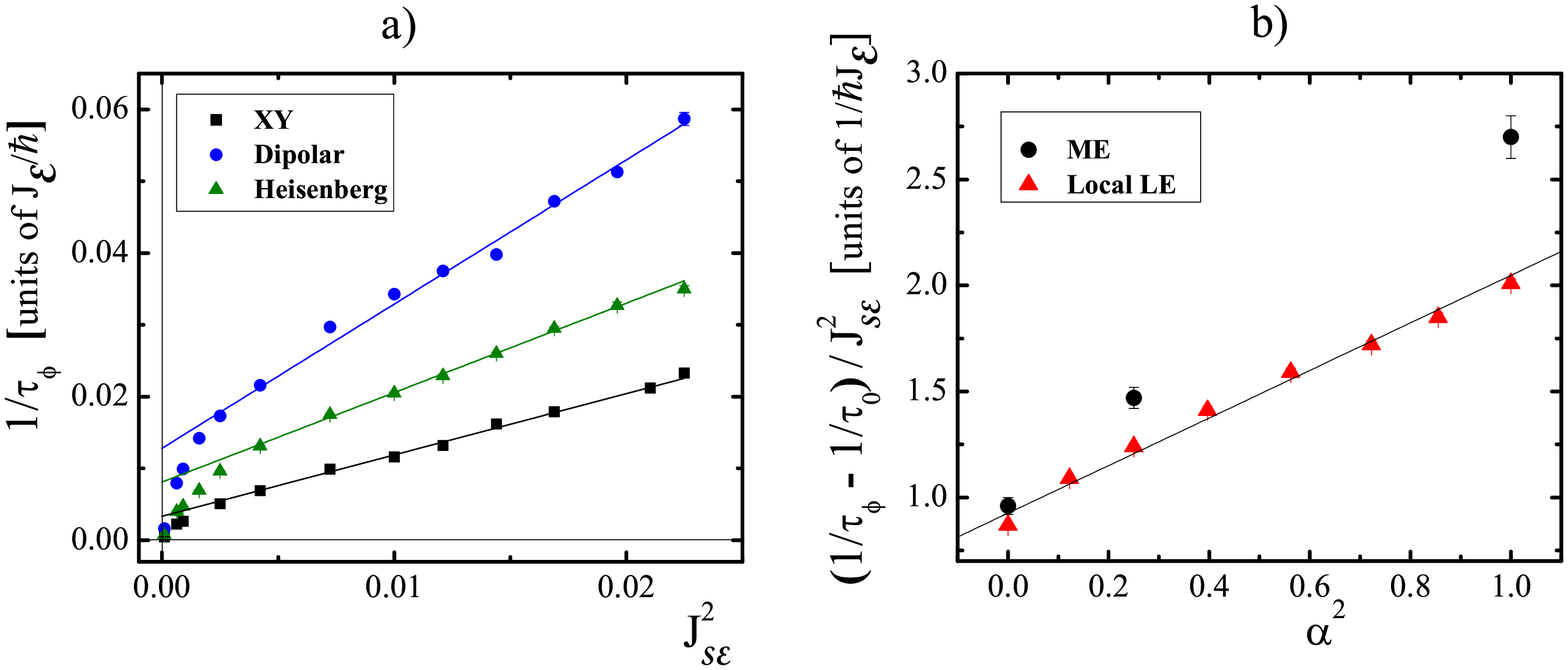}\\
      \caption{(Color Online). \textbf{a}) Decay rates for the NMR relevant interchain couplings, in units of $J_{\mathcal{E}}^{{}}/\hbar$.
The slopes and offsets depend on the value of $\alpha$, i.e.
on the relative weight between the Ising (dephasing) and $XY$ (polarization
transfer) contributions to the interchain coupling. \textbf{b}) Scaled rates showing the additivity of
the XY and Ising contributions to the FGR rate as a function of $\alpha^{2}$. For comparison, the rates obtained from Mesoscopic
Echo degradation data obtained in Ref. \cite{Alvarez-ME} are also plotted.}%
    \label{Fig: DecayRate}%
\end{figure*}

In order to assess the exponential regime, we plot the characteristic rate
$1/\tau_{\phi}$ in Fig. \ref{Fig: DecayRate}-\textbf{a}, as a function of
$J_{\mathcal{SE}}^{2}$ in units of $J_{\mathcal{E}}^{{}}/\hbar$. This quantity
is appropriate to verify the FGR (Eq. \ref{fgr1}), as long as $J_{\mathcal{SE}%
}^{2}$ is the typical scale for the second moment of the $\mathcal{S}%
$-$\mathcal{E}$ interaction and $1/J_{\mathcal{E}}^{{}}$ that of $N_{1}%
$, the DDCS.
We point out that rate calculations with longer rings and chains are seen to give similar values as long as the above parameters are kept the same. Such independence on $m$ resembles to the SP and LE for a single spin interacting with chains of different lengths \cite{elenacpl,pablo-axel}. This is indicative that we are reaching the limit were the DDCS becomes dense enough to manifest the sum rule intrinsic to the FGR\cite{ElenaPhysicaB}. Thus, we restrict our analysis to cases with $m=5$.
Although several forms of the interchain Hamiltonians were
considered by varying the anisotropy parameter $\alpha$ , we show only those
relevant to NMR experiments: $XY$\ ($\alpha=0$), isotropic ($\alpha=-\frac
{1}{2}$), and truncated dipolar interaction ($\alpha=1$). We observe that the
boundary conditions play a non-trivial role \cite{danieli-CPL2004}. For the
case of open boundary conditions (Fig. \ref{Fig: system1}-\textbf{a}), some
oscillations appear mounted on a decay which also depend on the parity of
$m,$ the number of spins in each chain. Here, we present only the results using closed
boundary conditions (rings) where these effects are much weaker.%

Note that the Loschmidt Echo allows one to explore a range of very weak
perturbations yielding $\tau_{\phi}$ from the regime of exponential decay.
Such a range was inaccessible in the interference contrast method
\cite{Alvarez-ME}. We observe that the rates in Fig. \ref{Fig: DecayRate}-a
start from zero and have a rapid increase with the $J_{\mathcal{SE}}$ perturbation. After a
small perturbation threshold the rates slow down to a linear dependence on the second moment of the perturbation. This confirms the validity of Eq. \ref{fgr1} for this range. The
linear fit is shifted by an offset, $1/\tau_{0}$, which seems to depend
on the nature of the $\mathcal{SE}$ interaction as it becomes larger for those
perturbations with bigger Ising components.

Fig. \ref{Fig: DecayRate}-\textbf{b }shows the FGR contribution to the
scaled decoherence rates $[1/\tau_{\phi}-1/\tau_{0}]/J_{\mathcal{SE}}^{2}$ as a function of interaction anisotropy $\alpha^{2}$.
There, we also include the
rates derived from the ME attenuation \cite{Alvarez-ME}. From the slope in
that plot we derive the contributions to the global decay rate $1/\tau_{\phi}%
$, arising from $XY$ and Ising processes in the interchain interaction:%

\begin{equation}
\frac{1}{\tau_{\phi}}=\frac{1}{\tau_{0}}+\frac{1}{\tau_{\phi}^{XY}}+\frac
{1}{\tau_{\phi}^{ZZ}}.\label{tauphi}%
\end{equation}
For the Loschmidt Echo results:%
\begin{equation}
LE:~~~\frac{1}{\tau_{\phi}^{XY}}=(0.92\pm0.04)\frac{J_{\mathcal{SE}}^{2}%
}{\hbar J_{\mathcal{E}}^{{}}},\label{tauxy}%
\end{equation}%
\begin{equation}
LE:~~~\frac{1}{\tau_{\phi}^{ZZ}}=(1.12\pm0.04)\alpha^{2}\frac{J_{\mathcal{SE}%
}^{2}}{\hbar J_{\mathcal{E}}^{{}}}.\label{tauzz}%
\end{equation}
In order to compare with the numerical results in Ref. \cite{Alvarez-ME}, and
translating its notation $-a/b\equiv2\alpha$, the rates contributing to the ME
degradation result:%

\begin{equation}
ME:~~~\frac{1}{\tau_{\phi}^{XY}}=(1.00\pm0.06)\frac{J_{\mathcal{SE}}^{2}%
}{\hbar J_{\mathcal{E}}^{{}}},\label{tauxy_meso}%
\end{equation}%
\begin{equation}
ME:~~~\frac{1}{\tau_{\phi}^{ZZ}}=(2.0\pm0.3)\alpha^{2}\frac{J_{\mathcal{SE}%
}^{2}}{\hbar J_{\mathcal{E}}^{{}}}.\label{tauzz_meso}%
\end{equation}
The most striking effect evinced by this comparison is that while the $XY$
contributions are essentially the same in both methods, the Ising contribution
(i.e. pure dephasing) in the ME\ (Eq. \ref{tauzz_meso}) is almost twice the
value obtained from the local LE. In the following section we analyze the
origin of such result.

\subsection{Analysis of the Decoherence Regimes \label{Sec: Deco Rate Analysis}}

First, we would like to address the applicability of a ``\textit{SP picture}"
for the LE in the present case, as it was proposed in Section
\ref{Sec: QD of Spin Chain} in connection with previous works. We verified
that the short time decay is naturally ruled by the second moment of the
$\mathcal{S}$-$\mathcal{E}$ interaction $\sigma^{2}=[J_{\mathcal{SE}}^{{}%
}/2]^{2}$. We also checked the estimation for the spreading time $t_{s}%
\simeq\hbar/J_{\mathcal{E}}^{{}}$, \textit{i.e.} the transition from the
quadratic short-time to the exponential regime. We found that such prediction
is in agreement with the observed end of the quadratic decay.

In our spin problem, the LE exponential regime needs a further delay beyond
$t_{s}$ to become well defined. This indicates that the DDCS, is not
immediately defined but $J_{\mathcal{SE}}^{}$\ may have a role in its
stabilization. Indeed, it is needed a back action of $J_{\mathcal{SE}}^{{}}%
$\ to break the strong degeneracy present in the unperturbed integrable
$\mathcal{E}$, a finite $XY$ chain itself. Only then, one has the sufficiently dense
spectrum required for the FGR to apply. Once the whole Hilbert space is made accessible by the perturbation, as many as $\tbinom{2m}{m}$ states become coupled by an interaction that conserves spin projection.

The interpolative Eq. \ref{Gauss-FGR-interpolation} approximates
satisfactorily the short-time decay, but it is not as well suited for
intermediate times and certainly can not be applied to the saturation regime. These
deviations can be, in principle, related to a failure in the SP picture to
describe the local LE dynamics. The physics underlying Eq.
\ref{Gauss-FGR-interpolation} is based on the assumption of a sufficiently
large number of particles and a well defined $N_{1}$. Also, one should be
aware that our spin case is strictly finite, indeed quite small, and therefore
the polarization cannot relax to zero. The observed asymptotic steady
magnetization $1/2m$ can be identified with an ergodic behavior of the excitation described by the local LE
dynamics. Since this ergodicity is not present in the unperturbed system, it
should emerge as a consequence of the $\mathcal{SE}$ interaction.

It is important to notice that neither the rate obtained nor $\sigma
^{2}$ depend on the total number of spins or on the number of
spins in $\mathcal{S}$. The reason for such independence is related to the initial non-equilibrium condition stated in Eq.
\ref{exitacion-inicial}, which is a well localized excitation that maintains
its character when it propagates through an $XY$ chain. It turns out that such
localized excitation behaves much like a single particle and, in some sense, almost classically.
This remains true at least when the LE is in the exponential regime. We have seen that the independence on the number
of spins does not hold if one considers an initial state built-in as an
arbitrary superposition. However, the investigation of this issue goes beyond the
scope of the present study.

One should notice that splitting the FGR contribution to the decoherence rate
into two well separated terms, a strategy also exploited in Ref.
\cite{Alvarez-ME}, evidences the additivity of the $XY$ and Ising processes.
Each of them is associated with a different term in the
interaction coupling\ $\hat{V}_{\mathcal{SE}}^{{}}$ and require a different DDCS.
These properties are indeed manifested in Fig. \ref{Fig: DecayRate}, that shows the linear dependence of $(1/\tau_{\phi}-1/\tau_{0})/J_{\mathcal{SE}}^{2}$\ on the relative weight
$\alpha^{2}$.

Comparing the rates obtained through the LE with those computed by the ME
degradation \cite{Alvarez-ME}, we notice that the $XY$ rates are
equivalent and we can interpret such equivalence by using the mapping into a one-body
evolution in both $\mathcal{S}$ and $\mathcal{E}$. Even when the one-body
picture is not rigorously valid when the interaction $\hat{V}_{\mathcal{SE}%
}^{{}}$ is turned on, it provides some physical insights \cite{pablo-axel} that apply to more complex cases. Accordingly, the dynamics along the chains is only weakly
affected by the tunnelling processes, (i.e. in a single particle picture, the
kinetic energy along the chains commutes with that along the interchain
direction). Thus, the rate $1/\tau_{\phi}^{XY}$ should coincide with the
interchain tunnelling rate and we expect that it should not be affected by a
time-reversal procedure within $\mathcal{S}$.

\begin{figure}
    \centering
     \includegraphics[width=0.5\textwidth]{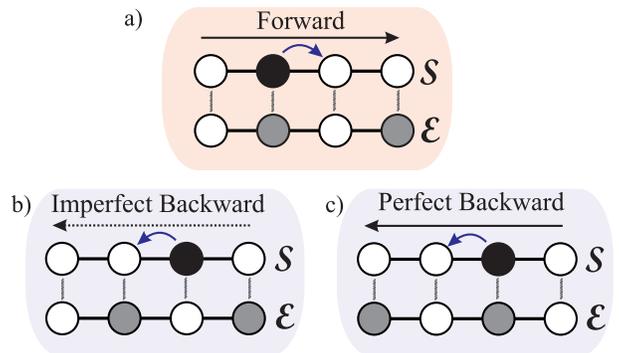}\\
      \caption{Spin propagation in presence of an Ising interchain interaction, represented as a single fermion moving through a binary alloy landscape. A filled circle is a spin up or a fermion, and an empty circle represents a spin down or a hole. All the processes are considered in the short time
scale given by a single Trotter time step.
 \textbf{a}) Forward dynamics within the upper chain $\mathcal{S}$, in presence of the "random site energies" produced by the Ising interaction with the lower chain $\mathcal{E}$.
\textbf{b}) If $\mathcal{E}$ remains frozen, the backward evolution in $\mathcal{S}$ is imperfect because the signs of  the \textquotedblleft site energies\textquotedblright\ are not inverted.
 \textbf{c})A particular evolution of $\mathcal{E}$ provides a perfectly reversed dynamics for $\mathcal{S}$. In this case, the local portion of potential landscape, that determined the excitation's forward evolution, is reversed.
}%
    \label{energyfluctuation}%
\end{figure}

In contrast to the simple decay associated to the $XY$ component of the
$\mathcal{S}$-$\mathcal{E}$ interaction, the Ising part causes energy fluctuations that induce a
diffusion-like process within the system. This tends to blur out the dynamical recurrences (i.e. MEs). The
smaller decoherence rate observed from the LE indicates that the time reversal
procedure is at least partially effective to reverse such processes. In fact,
the rates for the pure Ising interaction observed here from the LE are about
half of those obtained from the ME \cite{Alvarez-ME}. This means that the ME
attenuation overestimates the phase degradation induced by the environment. As
stated before, the displacement of a spin excitation in presence of a quenched
spin environment plays the role of an Anderson disorder degrading the wave
packet dynamics that produces the ME. Such degradation is computed as decoherence. The
reversal of the internal $XY$ interaction would not be able to reverse such
disorder. However, if the environment has an inner dynamics in a time scale comparable to that of the system ($J_{\mathcal{S}}^{}\simeq{}J_{\mathcal{E}}^{}$), there are particular fluctuations that allow for a perfect reversal of the system dynamics.
Indeed, this occurs when both the hopping and the site energy signs are inverted. Such specific
$\mathcal{E}$-fluctuations are those needed to unravel the phase shifts
produced during the forward evolution. This argument, which relies on a single particle picture and equivalent time scales for the local dynamics in $\mathcal{S}$ and $\mathcal{E}$, is represented in Fig. \ref{energyfluctuation}.
Thus, one may safely say that in the presence of a fluctuating environment, the LE is able to reconstruct the phases in an appreciable fraction of the local configurations.

\section{CONCLUSIONS\label{Sec: Conclusions}}

We presented an evaluation of decoherence for a spin chain in realistic
many-body scenarios. As specific realizations for structured environments, we
used a second spin chain weakly coupled to the first. This system-environment
interaction ranges from pure $XY$ to truncated dipolar, passing through the
isotropic Heisenberg interaction. In order to evaluate decoherence, we
resorted to the Loschmidt Echo, which here is the local polarization recovered
after an imperfect time reversal procedure. The attenuation of such echoes
yields a first hand estimation of the decoherence rate, without any
\textit{ad-hoc} assumptions about spectral functions
\cite{chakraLegget84,weiss-book} or stochastic noise operators
\cite{serva-JSP1994,pascazio-jourSuper99}\textbf{. }Our computational
technique involves a Trotter Suzuki decomposition assisted by a recently
developed algorithm that relies on quantum parallelism to evaluate local
observables \cite{Alv-parallelism}.

In the present many-body context, the decoherence rate is separated in two
contributions, both of which scale with $J_{\mathcal{SE}}^{2}$ and
$1/J_{\mathcal{E}}$, i.e. as a Fermi golden rule. The rates obtained here do not
depend on the number $m$\ of spins. This is strongly related to the localized
initial condition but such independence does not hold if the initial condition
is a superposition state. It is also indicative of a specific sum rule
relating the \textit{local} second moment of the interaction and the DDCS, which
in this many-body case coincides with that resulting from a single particle picture.

In the adopted model, the close connection between the Loschmidt Echo and the
Survival Probability allows for a comparison with the expectancies of the last
magnitude. In particular, we assessed an interpolative formula \cite{Flambaum2001} that matches
the initial quadratic decay with the exponential regime. Despite of the
qualitative agreement, the Loschmidt Echo evidences a richer and more complex
dynamical behavior than those predictions. For instance, the breakdown of
single-particle degeneracies due to the $\mathcal{S}$-$\mathcal{E}$
interaction and the appearance of an ergodic regime, manifested as a steady
saturation, are now clearly shown in the numerical results.

The numerical studies performed here confirm that the Loschmidt Echo is
better to quantify decoherence than the standard analysis based on interference
degradation, as it recovers information that escaped such analysis
\cite{Alvarez-ME}. Additionally, by compensating the intrinsic dynamical
interferences of the system through time reversal, the LE gets rid of most of
the trivial part in the $\mathcal{S}$ dynamics that conceals the decoherence
effects. Thus the LE provides a smooth and continuous access to characterize
decoherence processes.

\section{Acknowledgements}

We acknowledge Fernando Pastawski for critical comments on an initial version
of this manuscript as well as Gonzalo Usaj and Greg Boutis for lively
discussions. This work was performed with the financial support from CONICET,
ANPCyT, SeCyT-UNC, MinCyT-Cor, and NVIDIA professor partnership program.

\begin{widetext}
\appendix

\section{The spin autocorrelation function.\label{Sec: Apendice2}}

We show here equivalent expressions for the spin autocorrelation
expressed in Eq. \ref{autocorrelation}.
The autocorrelation $P_{1,1}(t)$ gives the local polarization at
time $t$ along the $z$ direction in site $1$ provided that at time $t=0$ the
system was in its thermal equilibrium state plus a local excitation in site
$1$. In order to show this assertion explicitly, we rewrite the numerator in
Eq. \ref{autocorrelation}:%

\begin{equation}
\left\langle \Psi_{eq}^{{}}\right\vert \hat{S}_{1}^{z}(t)\hat{S}_{1}%
^{z}(0)\left\vert \Psi_{eq}^{{}}\right\rangle =\left\langle \Psi_{eq}^{{}%
}\right\vert \left(  \hat{S}_{1}^{+}(t)\hat{S}_{1}^{-}(t)-\frac{1}{2}\right)
\left(  \hat{S}_{1}^{+}(0)\hat{S}_{1}^{-}(0)-\frac{1}{2}\right)  \left\vert
\Psi_{eq}^{{}}\right\rangle ,\label{a1}%
\end{equation}

where we have used the identity $\hat{S}_{{}}^{z}=\hat{S}_{{}}^{+}\hat{S}_{{}%
}^{-}-\frac{1}{2}$, which is valid for spins $1/2$. The $\mathcal{S}$ label has been dropped for simplicity. Then,%

\begin{equation}
\left\langle \Psi_{eq}^{{}}\right\vert \hat{S}_{1}^{z}(t)\hat{S}_{1}%
^{z}(0)\left\vert \Psi_{eq}^{{}}\right\rangle =\left\langle \Psi_{eq}^{{}%
}\right\vert \hat{S}_{1}^{+}(t)\hat{S}_{1}^{-}(t)\hat{S}_{1}^{+}(0)\hat{S}%
_{1}^{-}(0)\left\vert \Psi_{eq}^{{}}\right\rangle -\frac{1}{4},\label{a2}%
\end{equation}

where we have rearranged the terms and used the high temperature hypothesis
for $\left\vert \Psi_{eq}^{{}}\right\rangle $ explicitly, i.e. $\left\langle
\Psi_{eq}^{{}}\right\vert \hat{S}_{1}^{z}(t)\left\vert \Psi_{eq}^{{}%
}\right\rangle =\left\langle \Psi_{eq}^{{}}\right\vert \hat{S}_{1}%
^{z}(0)\left\vert \Psi_{eq}^{{}}\right\rangle =0$. It is crucial now to
identify the meaning of the expectation value $\left\langle \Psi_{eq}^{{}%
}\right\vert \cdots\left\vert \Psi_{eq}^{{}}\right\rangle $ as a trace over
the basis states (see Eq. (3) in Ref. \cite{danieli-CPL2004}). Hence, the
invariance of the trace under cyclic permutation leads to:%
\begin{equation}
\left\langle \Psi_{eq}^{{}}\right\vert \hat{S}_{1}^{z}(t)\hat{S}_{1}%
^{z}(0)\left\vert \Psi_{eq}^{{}}\right\rangle =\left\langle \Psi_{eq}^{{}%
}\right\vert \hat{S}_{1}^{-}(0)\hat{S}_{1}^{+}(t)\hat{S}_{1}^{-}(t)\hat{S}%
_{1}^{+}(0)\left\vert \Psi_{eq}^{{}}\right\rangle -\frac{1}{4},\label{a3}%
\end{equation}%
and one identifies the non-equilibrium initial condition $\hat{S}%
_{1}^{+}(0)\left\vert \Psi_{eq}^{{}}\right\rangle $, which after normalization
yields exactly Eq. \ref{exitacion-inicial}:%
\begin{equation}
\left\vert \Psi_{ne}^{{}}\right\rangle =\frac{\hat{S}_{\mathcal{S},1}%
^{+}\left\vert \Psi_{eq}^{{}}\right\rangle }{\left\vert \left\langle \Psi
_{eq}^{{}}\right\vert \hat{S}_{\mathcal{S},1}^{-}\hat{S}_{\mathcal{S},1}%
^{+}\left\vert \Psi_{eq}^{{}}\right\rangle \right\vert ^{1/2}}.\label{a5}%
\end{equation}
Accordingly, the numerator can be written as:%
\begin{align}
\left\langle \Psi_{eq}^{{}}\right\vert \hat{S}_{1}^{z}(t)\hat{S}_{1}%
^{z}(0)\left\vert \Psi_{eq}^{{}}\right\rangle  & =\left\vert \left\langle
\Psi_{eq}^{{}}\right\vert \hat{S}_{\mathcal{S},1}^{-}\hat{S}_{\mathcal{S}%
,1}^{+}\left\vert \Psi_{eq}^{{}}\right\rangle \right\vert \left\langle
\Psi_{neq}^{{}}\right\vert \hat{S}_{1}^{+}(t)\hat{S}_{1}^{-}(t)\left\vert
\Psi_{neq}^{{}}\right\rangle -\frac{1}{4},\nonumber\\
& =\frac{1}{2}\left\langle \Psi_{ne}^{{}}\right\vert \hat{S}_{1}^{+}(t)\hat
{S}_{1}^{-}(t)\left\vert \Psi_{ne}^{{}}\right\rangle -\frac{1}{4}.\label{a6}%
\end{align}

The denominator in Eq. \ref{autocorrelation} is $\left\langle \Psi_{eq}^{{}%
}\right\vert \hat{S}_{1}^{z}(0)\hat{S}_{1}^{z}(0)\left\vert \Psi_{eq}^{{}%
}\right\rangle \equiv\frac{1}{4}$. Therefore,%
\begin{equation}
P_{1,1}(t)=2\left\langle \Psi_{ne}^{{}}\right\vert \hat{S}_{1}^{+}(t)\hat
{S}_{1}^{-}(t)\left\vert \Psi_{ne}^{{}}\right\rangle -1.\label{a4}%
\end{equation}

In the Schr\"{o}dinger picture it yields:%
\begin{align}
P_{1,1}(t)  & =2\left\langle \Psi_{ne}^{{}}\right\vert e^{\mathrm{i}\hat
{H}_{total}^{{}}t/\hbar}\hat{S}_{1}^{+}\hat{S}_{1}^{-}e^{-\mathrm{i}\hat
{H}_{total}^{{}}t/\hbar}\left\vert \Psi_{ne}^{{}}\right\rangle -1,\label{a7}\\
P_{1,1}(t)  & =2\left\langle \Psi_{ne}^{{}}\right\vert e^{\mathrm{i}\hat
{H}_{total}^{{}}t/\hbar}\left(  \hat{S}_{1}^{+}\hat{S}_{1}^{-}-\frac{1}%
{2}\right)  e^{-\mathrm{i}\hat{H}_{total}^{{}}t/\hbar}\left\vert \Psi_{ne}%
^{{}}\right\rangle ,\label{a8}\\
P_{1,1}(t)  & =2\left\langle \Psi_{ne}^{{}}\right\vert e^{\mathrm{i}\hat
{H}_{total}^{{}}t/\hbar}\hat{S}_{1}^{z}e^{-\mathrm{i}\hat{H}_{total}^{{}%
}t/\hbar}\left\vert \Psi_{ne}^{{}}\right\rangle .\label{a9}%
\end{align}%

Thus, the autocorrelation $P_{1,1}(t)$ written in the form of
Eq. \ref{a9} is, in fact, the expectation value of the $\hat{S}_{1}^{z}$
operator over the evolved non-equilibrium state $e^{-\mathrm{i}\hat{H}%
_{total}^{{}}t/\hbar}\left\vert \Psi_{ne}^{{}}\right\rangle $. We note that an equivalent reasoning can be performed in the case of the local
Loschmidt Echo in Eq. \ref{loca-Many-Body Loschmidt} just replacing the forward propagator  $e^{-\mathrm{i}\hat{H}
_{total}^{{}}t/\hbar}$ by the Loschmidt Echo propagator $e^{-\tfrac{\mathrm{i}}{\hbar
}(-\hat{H}_{\mathcal{S}}^{{}}+\hat{\Sigma})t_{R}} e^{-\tfrac{\mathrm{i}
}{\hbar}(\hat{H}_{\mathcal{S}}^{{}}+\hat{\Sigma})t_{R}}$.

\section{Numerical Simulations with Trotter-Suzuki evolutions.
\label{Sec: Apendice}}

The computation of the time dependent observable in Eq. \ref{autocorrelation}
requires to evolve each of the states participating in the ensemble. However,
its implementation for large systems, e.g. by means of a $4^{th}$ order
Trotter-Suzuki decomposition \cite{trotter}, soon becomes quite expensive in
computational resources. Instead, the local nature of the excitation allows
the employment of an algorithm exploiting the quantum parallelism
\cite{Alv-parallelism} to give an exponential reduction of computational
efforts. The evolution of a few pure states is enough to describe the ensemble
averaged excitation dynamics. Thus, a typical initial state representing the
whole ensemble has the form:%
\begin{equation}
\left\vert \Psi\right\rangle =\left\vert \uparrow_{1}\right\rangle
\otimes\left\{
{\displaystyle\sum\limits_{r=1}^{2^{2m-1}}}
\frac{1}{\sqrt{2^{2m-1}}}e^{\mathrm{i}\varphi_{r}}\text{\ }\left\vert
\beta_{r}\right\rangle \right\}  ,\text{ \ \ \ }\varphi_{r}^{{}}=\text{random
phase,}\label{initial3}%
\end{equation}
which exploits the quantum superposition over the whole spin set. Typically, a
few of these entangled states is enough to get rid of statistical noise and
obtain local physical observables with good accuracy. This highly efficient technique for
spin-ensemble calculations is enhanced by the parallelization enabled by its
implementation on a General Purpose Graphical Processing Unit (GPGPU).
\end{widetext}

%

\end{document}